# Taming the paper pile at home: Adopting Personal Electronic Records

**Matt Balogh**

University of New England, Armidale, NSW, Australia

https://doi.org/10.48550/arXiv.2204.13282

## Abstract

Research has found that if respondents do not manage their personal records such as bills, receipts and tax-related documents efficiently, they risk not being able to re-find them when needed, resulting in significant problems.  A significant gap in understanding and addressing this problem stems from a lack of knowledge of the format of these records, particularly in the context of the COVID-19 pandemic, that may have caused an increase in managing personal records in an electronic format, rather than by hardcopy.  This paper provides results of quantitative research conducted in 2018, thereby providing a valuable benchmark for future research on the same and related topics.  This measurement was achieved by means of an online survey distributed via social media amongst 205 respondents. The results revealed that nearly all respondents (97%) retained at least some records, and more than 80% kept some of those records in an electronic format, particularly travel reservations and payslips.   Conversely, only 10% of respondents kept receipts and warranties for appliances or medical records in an electronic format.  The reason for these differences in propensity of keeping various records in an electronic or hardcopy format will require further research.

**Keywords:** Personal Electronic Records Management, (PERM), Personal information management (PIM), Records management, email, folders

## Introduction

Prior research (Balogh, Kennan, Billingsley, & Paul, 2022) has found a diversity of ways in which respondents manage their personal records.  Research has found examples of bills not being paid on time, registration and insurance payments being overlooked, and identified difficulties in re-finding information when required, such as for taxation reporting purposes. An improved understanding of the behaviours and the respondents' needs with regards to their personal electronic records aids the development of systems which assist respondents in their personal records management. Despite extensive research regarding personal information management PIM (Dinneen & Julien, 2020), in practice the study of PIM has predominantly focussed on workplaces and particularly academic institutions (Balogh et al., 2022; Buttfield-Addison, 2014).  In this paper the word 'records' is used to describe the mixture of items that respondents retain at home, such as emails, bills and notes – as distinguished from workplace information and documents. The word 'records' also accommodates the diversity of formats in which information and documentation may manifest. For instance, some records may comprise just a few lines of text within an email or file, while other records may consist of a set of files (Finnell, 2011; Kim, 2013; McKemmish, 1996; Yeo, 2018).

Over the past thirty years, it has become possible to receive and maintain electronic versions of personal records, many of which do not have a hardcopy version – the electronic version is the sole record of the dealing (Alba, 2013; Balogh et al., 2022; Zacklad, 2019). Yet research has found that respondents are not satisfied with how they manage their personal information (Alon & Nachmias, 2020b), eliciting language such as '*anxious*', '*frustrated*' and even '*desperation*' (Alon & Nachmias, 2020a). Furthermore, as Kalms observed:

> `*While much is known about the modern household as a[n]… adopter of technologies, little is known about how the household… manages the information it receives each day*'
> (2008, p. 2).

Consequently, there is a need to better understand how respondents retain their personal electronic records and particularly the proportion of those documents that are kept in hardcopy or electronic formats (or both formats) in order to develop systems to make it easier to manage and re-find the records that respondents need when required (Balogh et al., 2022).

## Related Research

The study of personal records management at home transcends: the field of personal information management (PIM), which tends to study information management in the workplace (Balogh et al., 2022), records management; which can include a wide variety of records that respondents deal with at home (Hobbs, 2001; McKemmish, 1996) and digital archiving, which studies the long term maintenance of personal electronic records (Kim, 2013; Kirk et al., 2010). Yet little is known of the everyday practices and formats of personal records, such as bills, insurance documents or tax records that are essential for everyday life.

Smith provided a list of thirty-six common documents, of which the personal documents that respondents might keep at home included marriage and birth certificates, medical records, bank or pension records, and payslips (Smith, 2011, p. 3). Yet every month respondents are also receiving household bills, booking events or vacations online and making online purchases. What proportion of these dealings are electronic versus hardcopy? And how are these records maintained, if at all? A study in 1996 described the use of the email in-box in lieu of a to-do folder, observing that users often forgot to look at their to-do list, typically abandoning a to-do list or folder as an effective strategy (Whittaker & Sidner, 1996, p. 279). Conversely, the study cites only one effective to-do list strategy used by a participant, wherein their 'to-do' items appeared at the top of their email in-box. The effectiveness of the in-box as a to-do list was characterised by a research participant who said that they '*lived in the in-box*' (Whittaker & Sidner, 1996, p. 281). Ducheneaut and Bellotti go well beyond the description of email as '*the killer application of the Internet*' with the suggestion that '*e-mail is the serial-killer application!*' (Ducheneaut & Bellotti, 2001, p. 37) – describing email as the de-facto personal information management tool, but often without the necessary features. It is estimated that over 300 billion emails will be sent in 2020, increasing by around 20 billion per annum (Clement, 2019) to 361 billion per annum in 2024 (Mason, 2021, p. iii).

In order to better understand the management of personal records at home, we need to first understand to what extent respondents receive and retain their personal records such as bills in hardcopy or electronic formats.

# Research method

A quantitative survey was conducted to address the question of the way in which respondents receive and retain personal records. The survey was conducted online using social media posts to invite respondents. The survey was open for three days in August 2018, with three reminder posts in order to maximise response rates. Hence the research was conducted approximately 18 months before the onset of the COVID-19 Pandemic in early 2020. The survey took approximately two minutes to complete, and this was communicated to respondents in order to encourage participation. Survey respondents were shown a list of document types. For each item, research respondents were asked if they kept these documents in an electronic format or hardcopy format. Respondents also had the option to respond that the document type was not applicable to them. Only one option was permitted for each form of record. The analysis was conducted using IBM SPSS. Comparative percentages by age group are only reported if statistically significant, calculated by means of a Z Test using the Bonferroni correction ([Napierala, 2012](); [Weisstein, 2004]()). All percentages have been based on all respondents so that they represent the proportion that kept records in a given format, and not the proportion of records in that category that were retained in each format.

# Results

### Profile of respondents

In total 205 respondents completed the survey. As the researcher was Australian, the social media distribution resulted in a majority of the respondents (160) from Australia, 23 from the United States, 8 from the United Kingdom, eight from Switzerland and six from a mixture of other countries. Using a midpoint of the age ranges, a mean age of respondents was calculated at 46 years. This compares to the Australian median age of 37 – however it should also be borne in mind that the survey excluded children under the age of 18. For the purposes of analysis by age, the sample was divided into two groups, those aged 18-44, comprising 103 respondents, and those age 45 and over, comprising 102 respondents.

### Mostly commonly retained records

The graph in Figure 1 shows the incidence of survey respondents who retained each type of record listed – combining both hardcopy and electronic formats of the records. The graph displays the response in the order of prevalence of each category of personal records. It shows that nearly everyone stated that they kept records relating to tax returns and tax assessments (97%), and the receipts and warranties for appliances (95%). The vast majority also kept manuals for appliances (85%), a similar proportion keep bank statements and eight-in-ten kept receipts for tax deductible expenses (81%) and payslips (79%).

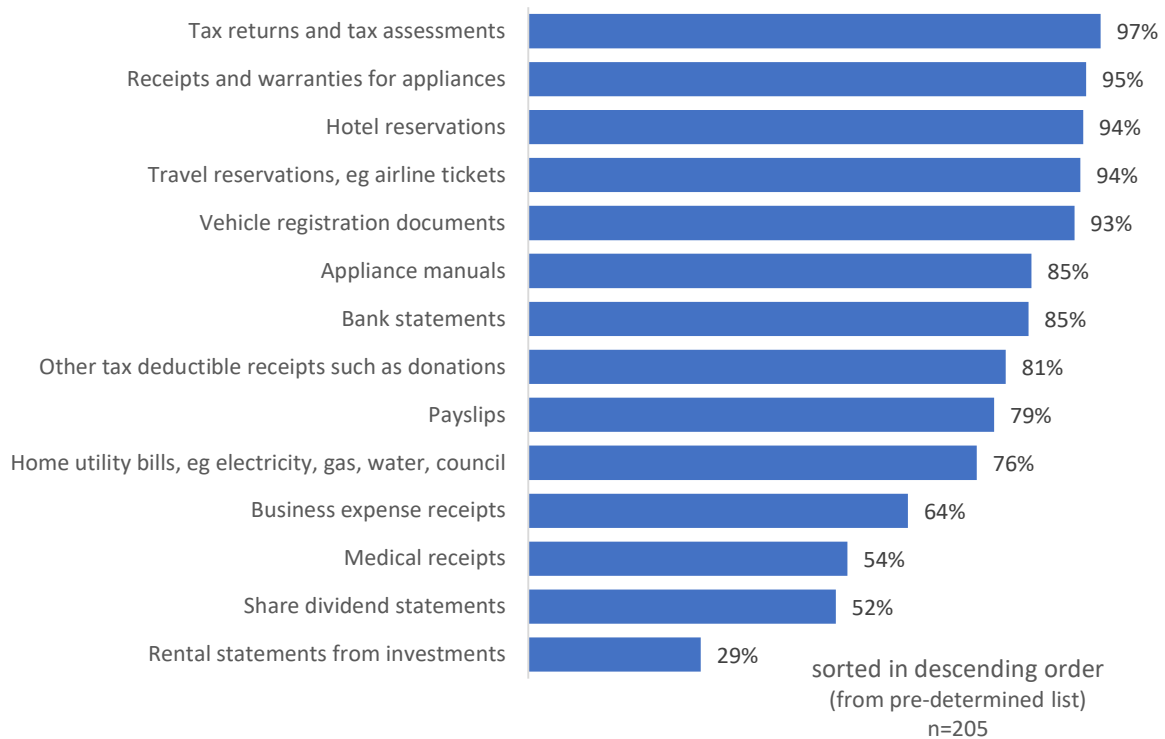

**Figure 1: Propensity to retain personal records**

Amongst the most prevalently retained records were travel and accommodation bookings, although further research would need to be conducted to determine whether these are kept after the travel event has occurred.

The graph in figure 1 shows that most respondents in this research retained all but one of the types of records asked about in the question. The one exception was rental receipts (from investment properties), which were only applicable to one-third (33%) of the respondents – however 88% of those who owned rental properties kept the statements that were provided to them.

**Additional document types and comments**
Respondents in this research were also asked if there were other kinds of records that they kept, or if they had other comments to make. In total, 20% of the respondents provided additional comments (40 out of 205). The two most commonly mentioned records retained that were not included in the list of items asked about in the question were wills and documents relating to children's schooling – particularly school reports.

Additional items mentioned by one or two respondents in the research were:

- Birth certificates
- Bank loan documents
- Credit card statements
- Credit card receipts
- Employment contracts
- Receipts for trade repairs
- Articles
- Recipes
- Health records and
- Records related to pet ownership & health.

**Propensity to retain records in electronic vs hardcopy format**

The propensity of survey respondents to retain personal records in an electronic or hardcopy format varied significantly for different types of records. For instance, travel documents, and particularly hotel bookings, were the most likely items to be retained electronically, by 80% of respondents in the case of hotel bookings, and 75% of travel bookings. At the other end of the scale, only 8% of respondents retained appliance manuals in an electronic format, compare to 78% that kept manuals in hardcopy.

In the following tables percentages comprise the proportion of survey respondents who retained the nominated personal records in an electronic or hardcopy format as a percentage of all respondents.

**Table 1 – Format of personal financial records**

| H% | Electronic | Hardcopy |
|---|---|---|
| Payslips/paystubs | 69% | 10% |
| Bank statements | 61% | 23% |
| Tax deductible receipts e.g. donations | 47% | 34% |
| Tax returns and tax assessments | 42% | 55% |

Table 1 shows that respondents to this survey reported a significantly higher incidence of retaining personal financial records such as payslips or pay stubs and bank statements in an electronic format, as opposed to a hardcopy format. With regards to payslips or pay stubs, respondents aged 18-44 had a higher incidence of keeping these records electronically (88%) compared to respondents aged 45 and over, of which 54% of thes respondents retained their payslips/pay stubs in an electronic format. Similarly, respondents aged 45 and over were less likely than those aged 18-44 to retain bank statements in an electronic foremat (59% compared to 67%). However, in the case of tax-related documents, the difference between the age groups was more significant – 53% of respondents aged 18-44 kept their tax records electronically (46 % in hardcopy), while amongst respondents aged 45 and over, only 34% kept their tax records electronically and 66% kept these records in a hardcopy format.

**Table 2 – Format of business and investment records**

| H% | Electronic | Hardcopy |
|---|---|---|
| Business expense receipts | 22% | 42% |
| Share dividend statements | 32% | 20% |
| Rental statements from investments | 20% | 9% |

Table 2 shows the propensity to retain certain forms of investment or business expense records in an electronic or hardcopy format. Percentages do not add up to 100% because not all the record types were applicable to all respondents – for instance, only 29% of respondents received rental investment statements. The results for these three categories were not analysed by age group as there were insufficient recipients of these forms of records amongst the 18-44 age group of respondents.

Amongst respondents who received statements of this type, in two thirds of cases, the statements were retained in an electronic format (20% of all respondents).

The results also show that business expense receipts were more likely to be retained in hardcopy than in an electronic format. Approximately one-third of respondents (32%) retained share dividend statements in an electronic format, compared to one-in-five (20%) who retained them in a hardcopy format.

**Table 3 Format of household records**

|  | H% Electronic | Hardcopy |
|---|---|---|
| Home utility bills, eg electricity, gas, water | 30% | 46% |
| Receipts and warranties for appliances | 10% | 85% |
| Appliance manuals | 8% | 78% |

The group of household record types shown in table 3 all had a significantly higher incidence of being retained in a hardcopy format rather than an electronic format. Nearly half (46%) of respondents reported retaining home utility bills in a hardcopy format, compared to less than a third (30%) who said they retained utility bills in an electronic format. This was one of a few records categories for which there was no statistically significant difference in responses by age group. With regards to receipts and warranties for domestic appliances, 85% of respondents retained these in hardcopy, compared to just one-in-ten (10%) that kept them in an electronic format. When asked about domestic appliance manuals, 78% of respondents said that they kept hardcopy versions of these, while only 8% retained electronic versions of their appliance manuals.

**Table 4 Format of motor vehicle records**

|  | H% Electronic | Hardcopy |
|---|---|---|
| Motor vehicle registration | 14% | 79% |
| Vehicle service records | 6% | 71% |

Table 4 displays the proportion of motor vehicle records retained in hardcopy or electronic format. In the case of motor vehicle registration documents, nearly eight-in-ten (79%) respondents reported keeping these in a hardcopy format, compared to 14% in electronic format. The propensity to keep vehicle registration documents electronically amongst respondents aged 18-44 was 19%, while only 9% of respondents aged 45 and over retained electronic versions of their vehicle registration documents, representing a moderate, but statistically significant difference by age group. With regards to vehicle service records, there was no significant difference by age group, with only 6% of respondents retaining service records in an electronic format, as compared to 71% who kept service records in hardcopy.

**Table 5 Format of medical receipts**

|  | H% Electronic | Hardcopy |
|---|---|---|
| Medical receipts | 10% | 44% |

Table 5 provides the survey results for the retention of medical receipts, showing that on-in-ten (10%) of respondents kept their medical receipts in an electronic format, while 44% kept medical receipts in a hardcopy format. This also reveals that 46% of respondents did not keep medical receipts in any format. There was no significant difference by age group for this category of personal records.

**Table 6 Format of travel documents**

| | H% Electronic | Hardcopy |
|---|---|---|
| Travel reservations, eg airline tickets | 75% | 19% |
| Hotel reservations | 80% | 14% |

Table 6 reports the proportion of respondents who maintained their travel reservations such as airline tickets, and specifically their hotel reservations in either an electronic or hardcopy format. Both of these two categories of personal records had a high incidence of being retained in an electronic format, 75% for travel reservations, and 80% for hotel bookings in particular. Approximately one-in-five (19%) respondents reported keeping travel reservations in hardcopy, and slightly less said that they kept hardcopies of hotel reservations (14%). While there was no statistically significant difference in the propensity to keep the broadly defined group of travel documents by age group, respondents aged 18-44 reported a measurably higher propensity to retain the sub-category of hotel reservations in an electronic format (87%) than respondents aged 45 and over (77%).

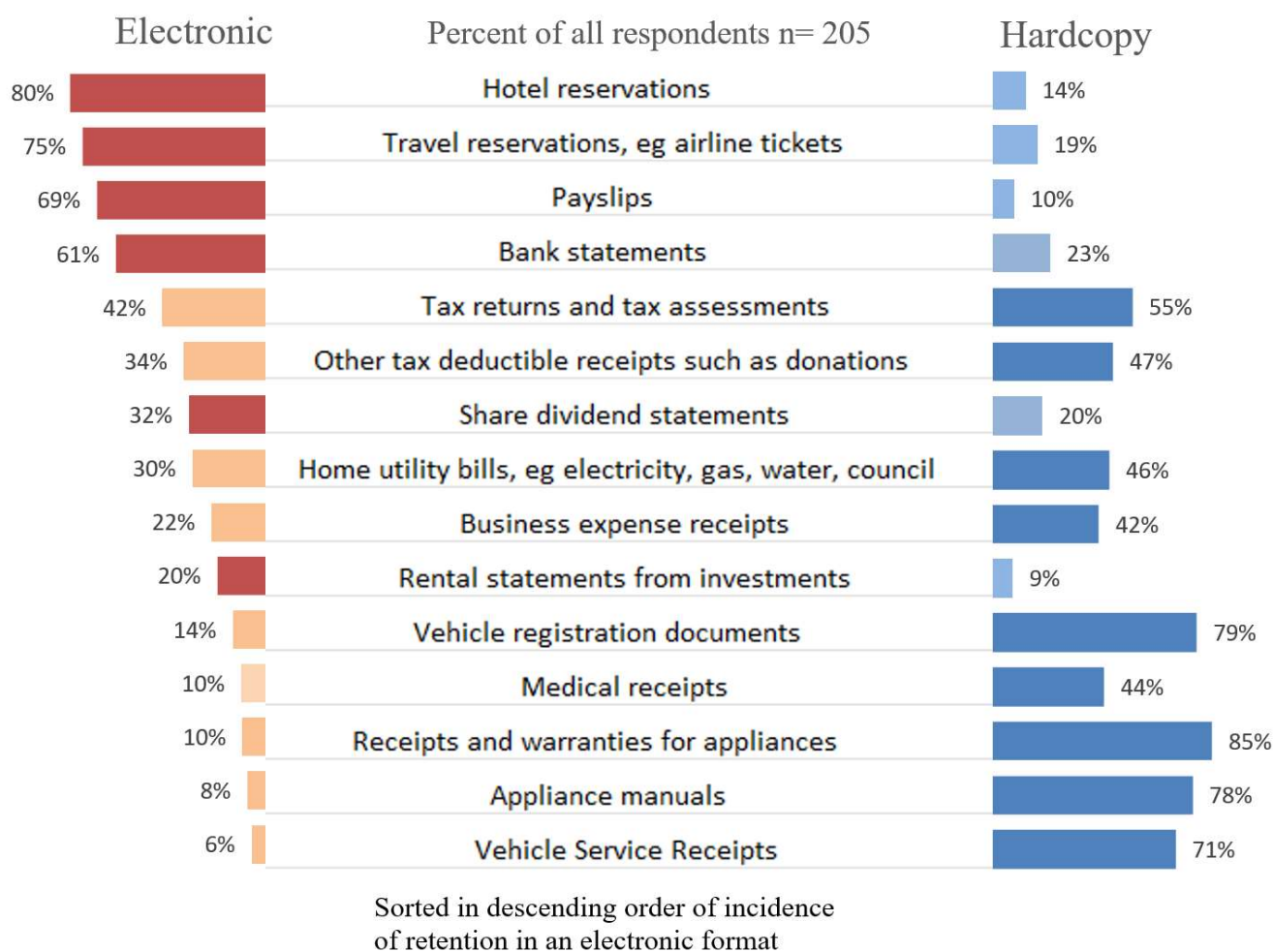

**Figure 2: Summary graph of personal records retention by format**

Figure 2 comprises a graph summarising the responses amongst all respondents. The column on the left shows the percentage of respondents who said that they retained each category of personal records in an electronic format. The column on the right shows the proportion of respondents who said that they retained each category of personal records in a hardcopy format. The darker shaded bars on the left highlight the categories for which more respondents said that they kept that category of records in an electronic format than the proportion that retained the same type of record in hardcopy. The darker shaded bars on the right highlight the categories for which more respondents said that they kept that category of records in a hardcopy format compared to an electronic format. As mentioned in the method section, percentages are based on all respondents and therefore also reflect the overall propensity to be keeping each type of category. This overall comparison shows that travel reservations, personal financial records, share dividend statements and rental income statements were more likely to be retained in an electronic format than as hardcopies. Records retained for tax reporting purposes were marginally more likely to be retained in a hardcopy format. Vehicle registration and service records, medical receipts and appliance warranties and manuals were predominantly retained in an electronic format.

**Respondent Comments**

Examination of the 40 comments added to the survey by some respondents indicated that some of the survey participants were using a higher proportion of paper records than they would prefer, for example:

> `Kids school reports and certificates and art work… yes drowning in paper.'

> `Credit card bills, I keep electronic copy. Any bills that I receive electronically are kept only electronically and those received as paper, I keep as paper. Prefer to receive electronic bills'

> `I keep the documents in the form they are provided to be, if I have the option of receiving them online then I do this'

> `I love the idea of keeping it all electronically & I usually never get around to scanning stuff that I get in hard copy. Whatever I can get electronically, stays that way'

However, interest in transitioning to electronic personal records can be impaired by a tendency to also retain hardcopy versions of the records as well, for example:

> `If documents are sent by email, I keep electronic but also print the document.'

> `I become completely overwhelmed with the task of filing paperwork! In many instances (most), I keep an electronic copy in addition to the hard copy.'

> `Also keep electronic copy of most of the above. Old habits die hard'

These comments were indicative of some internal conflict – on the one hand these respondents feel that retaining electronic versions of records would be beneficial, but on the other hand they are not ceasing to keep hardcopy versions of the same documents.

## Discussion

There are commonalities between the 36 common documents that respondents retain identified by Smith (2011) and the findings of this survey, such as tax related documents, payslips and banking records. However, some notable differences were travel reservations and receipts and warranties, which were amongst the most commonly retained documents according to the survey results.

The survey results show a distinct pattern with regards to whether personal records were retained in an electronic or hardcopy format. For instance, travel reservations, payslips or pay stubs and bank statements were more commonly retained in an electronic format rather than in hardcopy. Conversely, documents related to motor vehicles registration and vehicle servicing tended to be retained in a hardcopy format. For some categories, the propensity to retain records in one format or the other was statistically associated with the age of the respondent. For instance, the retention of pay slips in an electronic format was higher amongst respondents aged 18-44 (88%) compared to respondents aged 45 and over (54%). However, for other categories, such as household bills, the prevalence of retaining records in an electronic or hardcopy format was not statistically associated with the age of the respondents.

Additionally, there are two other important considerations with respect to these survey results. Firstly, technological progression: an increasing proportion of records are becoming available in an electronic format. For instance, personal driver licences became available in an electronic format in New South Wales, the most populous state of Australia, in November 2019. Inevitably, more services and transactions have been able to be completed digitally by those who choose to adopt that approach. Secondly, in early 2020, the COVID-19 pandemic brought about wide-spread lockdowns and avoidance of personal contact. In the developed world, digital location tracking and digital vaccination certificates became the norm. Online shopping in Australia grew by 16.6% in 2021 alone (Australia Post, 2022). Against this backdrop, we can expect that the adoption of electronic transactions and electronic keeping may well have increased substantially over 2020 to 2022. There is a clear opportunity to conduct further research on the incidence of personal records management and the propensity to maintain personal records in an electronic format at home in order to identify changes since the data for this research was collected in August 2018.

## Conclusion

This research provides a comparative measure of the incidence of retaining various personal records in an electronic format. The findings show that amongst these respondents, there are categories of personal records that were more frequently retained in an electronic format than in hardcopy – specifically travel reservations, payslips and bank statements. For other categories of personal records, the incidence of retention in hardcopy was only moderately greater than electronic format, such as tax related documents (52% hardcopy, 45% electronic), business expense receipts (47% hardcopy, 34% electronic) and household bills (46% hardcopy, 30% electronic). These findings show that all the survey respondents had at least some of their personal records in an electronic format in mid-2018. It is likely that the proportion of personal records retained electronically will have increased over the period 2020-2022 due to COVID-19. There is a significant opportunity for additional research on personal records and particularly personal electronic records management.

**Funding Information**

Matt Balogh is the recipient of an Australian Government Research Training Program Stipend Scholarship. Author states no other funding involved.

**Conflict of Interest**

Authors state no conflict of interest.